\documentclass[12pt]{article}
\usepackage{amsfonts,amssymb, amsmath}
\usepackage[english]{babel}

\DeclareMathAlphabet{\mathpzc}{OT1}{pzc}{m}{it}

\def\nn{\nonumber }
\def\bq{ \begin{equation} }
\def\eq{ \end{equation} }
\def\ben{ \begin{eqnarray} }
\def\en{ \end{eqnarray} }

\def\g{{\gamma}}

\newtheorem{re}{Remark}
\newenvironment{rem}{\begin{re} \rm }{\end{re}}
\begin{document}


\title{On the bi-Hamiltonian structure of Bogoyavlensky system on $so(4)$}
\author{A V Vershilov \\
\it\small
St.Petersburg State University, St.Petersburg, Russia\\
\it\small e--mail: alexander.vershilov@gmail.com \\
}

\date{}
\maketitle

\begin{abstract}
We discuss   bi-Hamiltonian structure for the Bogoyavlensky system on $so(4)$ with an additional integral of fourth order in momenta. An explicit procedure to find the variables of separation and  the separation relations  is considered in detail.
\end{abstract}


\vskip0.1truecm

\section{Introduction}
\setcounter{equation}{0}
We address the problem of the separation of variables for the Hamilton-Jacobi equation within the theoretical scheme of bi-hamiltonian geometry. The main aim is the construction of  variables of separation for the given integrable system without any additional information (Lax matrices, $r$-matrices, links with soliton equations etc.)

The paper is organized as follows.  In Section 2 we determine the Bogoyavlensky system on $so(4)$.  In Section 3, the necessary aspects of bi-hamiltonian geometry are briefly reviewed. Then, we discuss a possible application of these methods to calculation of the polynomial bi-hamiltonian structures for the given Bogoyavlensky system. In Section 4, the problem of finding variables of separation and corresponding separation relations is treated and solved.

All the  computations for the paper have been done by the computer algebra system {\sc Maple}.  It allows us to solve
overdetermined polynomial differential and algebraic systems of equations.  A major concern for any
operations performed by the program is the complexity of resulting expressions.  So, we can say that this note  belongs mostly to so called ``experimental'' or "computational" mathematical physics.

\section{The  Bogoyavlensky system on so(4)}
\setcounter{equation}{0}
Let $\g=(\g_1,\g_2,\g_3)$ and  $M=(M_1,M_2,M_3)$ be the two vectors of coordinates and momenta, respectively.
We postulate the following  Poisson bracket on  this six-dimensional phase space
\begin{equation} \label{p1}
\{M_{i},M_{j}\}=\varepsilon_{ijk}\,M_{k}, \qquad
\{M_{i},\g_{j}\}=\varepsilon_{ijk}\,\g_{k}, \qquad
\{\g_{i},\g_{j}\}=\varkappa^2 \, \varepsilon_{ijk}\,M_{k} .
\end{equation}
Here  $\varepsilon_{ijk}$ is the totally skew-symmetric and  $\varkappa$ is a parameter. It is well-known that any linear Poisson bracket is defined by an appropriate Lie algebra. The cases $\varkappa=0,$
$\varkappa^2>0$ and $\varkappa^2<0$ correspond to the Lie algebras $e(3)$, $so(4)$ and $so(3,1)$.

The Poisson bracket (\ref{p1}) has the two Casimir functions
\bq\label{caz}
C_1=M_1\g_1+M_2\g_2+M_3\g_3\,,\qquad
C_2=\varkappa^2(M_1^2+M_2^2+M_3^2)+\g_1^2+\g_2^2+\g_3^2\,.
\eq
Hence,  for the Liouville integrability of the equations of
motion only one additional integral functionally independent of the
Hamiltonian and the Casimir functions is necessary.

The nontrivial class of quadratic homogeneous Hamiltonians of the form
\begin{equation}
H=( M, \, \mathbf A  M)+( M, \, \mathbf B \g)+( \g, \, \mathbf C \g),
\label{HAMHOM}
\end{equation}
where $\mathbf A,\mathbf B$ and $\mathbf C$ are constant $3\times 3$-matrices,
has many important applications in the rigid body dynamics \cite{bm05}.

There are two classical integrable cases, one
found by Chaplygin and one by Goryachev, where the additional integral of motion is of fourth degree.
Namely,  at $\varkappa=0$  and $C_1=0$ integrals of motion are in the involution with respect to the Poisson bracket (\ref{p1})
 \ben\label{H1-0}
H_1&=& M_1^2+M_2^2+2M_3^2+c_1(\g_1^2-\g_2^2) \qquad c_1\in\mathbb R,\\
H_2&=&\left(M_1^2-M_2^2+c_1\g_3^2\right)^2+4M_1^2M_2^2\,.\nn
\en
It is so-called  Chaplygin system on the sphere \cite{chap}. The Lax matrices  and $r$-matrix formalism for this system have been obtained in \cite{kuzts89,ts02}, the corresponding bi-hamiltonian geometry  has been studied in \cite{ts07b,ts08c,ts10a}.

 Bogoyavlensky found pull-back of the Chaplygin system  to the   $so(4)$ algebra
\ben
H_1& =&
(1-\varkappa ^2a_1)M_1^2+(1-\varkappa ^2a_2)M_2^2+(2-\varkappa ^2a_2-\varkappa ^2a_1)M_3^2+(a_1-a_2)(\g_1^2-\g_2^2)\,,\nn\\
\label{bog-h}\\
H_2&=&
\Bigl((1-\varkappa ^2a_2)M_1^2-(1-\varkappa ^2a_1)M_2^2-(a_2-a_1)\g_3^2\Bigr)^2+4(1-\varkappa ^2a_2)(1-\varkappa ^2a_1)M_1^2M_2^2\,.\nn
\en
and proved that equations of motion can be integrated by means of elliptic functions \cite{bog91}.

\begin{rem}  In fact  we can found two integrable at $C_1=0$ systems on  $so(4)$  in the Bogoyavlensky book  \cite{bog91}.
The Hamilton function for the first system
\bq
\label{bog1b-H} \widetilde{H}_1= b_1 M_1^2+b_2 M_2^2+b_3 M_3^2+\frac{2
b_2-b_3}{\varkappa ^2}\, \g_1^2+\frac{2 b_1-b_3}{\varkappa ^2}\,
\g_2^2+\frac{b_2-b_3+b_1}{\varkappa ^2}\, \g_3^2
\eq
coincides  with the Hamiltonian $H_1$ (\ref{bog-h}) up to Casimir function
\[
\widetilde{H}_1=H_1+\frac{b_1+b_2-b_3}{\varkappa ^2}\,C_2\,,
\]
if we put
\[b_1-b_3=\varkappa ^2a_2-1,\qquad
b_2-b_3=\varkappa ^2a_1-1\,.
\]
The second  system with  the Hamilton  function
\bq\label{bog2-H}
\widehat{H}_1=c_1 M_1^2+c_2 M_2^2+c_3 M_3^2+\frac{1}{2\varkappa ^2}\Bigl((c_2+c_3)
\g_1^2+(c_1+c_3) \g_2^2+(c_1+c_2) \g_3^2\Bigr)
\eq
has the following additional integrals of motion
\[
K_i=\Bigl((c_i-c_k)\g_j^2+(c_j-c_i)\g_k^2+(c_j-c_k)\g_i^2\Bigr)^2+4(c_i-c_j)(c_i-c_k)\g_j^2\g_k^2,
\]
where $\{ijk\}$ is one of the possible cyclic permutations of subscripts $\{123\}$. It is easy to prove that
\[
\alpha K_1+\beta K_2+\gamma K_3=0, \quad\mbox{iff}\quad
\alpha+\beta+\gamma=0\,.
\]
According to \cite{tsbook}, Hamiltonians (\ref{bog1b-H} ) and (\ref{bog2-H}) are related by  the   Poisson map
\[M_1\to \dfrac{\g_1}{\varkappa },\quad \g_1\to \varkappa  M_1,\quad  M_2\to\frac{\g_2}{\varkappa },
\quad \g_2\to\varkappa  M_2,\quad M_3\to M_3,\quad \g_3\to \g_3,
\]
and change of parameters
\[c_1=(2b_2-b_3), \qquad c_2=(2b_1-b_3),\qquad  c_3 =b_3\,.\]
\end{rem}

\section{The bi-hamiltonian structure}
\setcounter{equation}{0}
In order to get variables of separation according to  general usage of bi-hamiltonian geometry firstly  we have to calculate the bi-hamiltonian structure for the given integrable system with integrals of motion $H_{1,2}$ (\ref{bog-h}) on the Poisson manifold $so(4)$ with the kinematic Poisson bivector $P$
 and the Casimir functions $C_{1,2}$  (\ref{caz}):
 \bq\label{p-1}
P=\left(
 \begin{array}{cccccc}
 0 & \varkappa^2 M_3 & -\varkappa^2 M_2 & 0 & \g_3 & -\g_2 \\
 - \varkappa^2 M_3 & 0 & \varkappa^2 M_1  & -\g_3 & 0 & \g_1 \\
  \varkappa^2 M_2 &  -\varkappa^2 M_1 & 0 & \g_2 & -\g_1 & 0 \\
 0 & \g_3 & -\g_2 & 0 & M_3 & -M_2 \\
 -\g_3 & 0 & \g_1 & -M_3 & 0 & M_1 \\
 \g_2 &-\g_1 & 0 &M_2 & -M_1 & 0 \\
 \end{array}
\right)\,,\qquad  PdC_{1,2}=0.
\eq
Following to \cite{ts07c,ts08,ts08b,ts09,ts10a} we are looking  for   solution $P'$ of the equations
\bq\label{m-eq12}
\{H_1,H_2\}'=\langle P'dH_1,dH_2\rangle=0,\qquad [P,P']=[P',P']=0,
\eq
where $[.,.]$ means the Schouten bracket.

Obviously enough, in their full generality equations (\ref{m-eq12}) are too difficult to be solved because it has infinitely many solutions \cite{ts07a,ts08b}. In order to get some particular solutions we will use the  additional assumption
\bq\label{m-eq3}
P'dC_{1,2}=0,
\eq
and polynomial in momenta $M$ \textit{ans\"{a}tze} for the components $P'_{ij}$ of the desired Poisson bivector  $P'=\sum P_{ij}\,\partial_i\wedge\partial _j$.

Substituting polynomial \textit{ans\"{a}tze} into the equations (\ref{m-eq12}-\ref{m-eq3}) and demanding that all the coefficients at powers of $M$ vanish one gets the over determined system of algebro-differential equations on functions of $x$ which can be easily solved in the modern computer algebra systems. The  computation was performed using the computer algebra system {\sc Maple} and, in contrast with the Chaplygin case \cite{ts10a},  these calculations were performed non-automatically with essential manual interaction.  In this way we get  a lot of real and complex  solutions, which will be classified and studied at a future date.

In this note  we will not consider a complete classification and  restrict ourselves by discussion of one  example only.
In order to describe this solution we  introduce some special notations. It is easy to see that at $C_1=0$ we can  rewrite kinematic bivector (\ref{p-1}) in the following form
\bq\label{l1}
P=\left(
  \begin{array}{cc}
   \varkappa^2 \partial_\g\Lambda  & \Lambda \\
   \\
    -\Lambda^\top & \partial_\g\Lambda
  \end{array}
\right)\,,\qquad
 \Lambda =\left(
                           \begin{array}{ccc}
                             0 & \g_3 & -\g_2 \\
                            - \g_3 & 0 & \g_1 \\
                             \g_2 & -\g_1 & 0 \\
                           \end{array}
                         \right)
\eq
where  antisymmetric matrix $\partial_\g \Lambda$  is defined by
\bq\label{dl1}
\Bigl(\partial_\g \Lambda \Bigr)_{ij}=\sum_{k=1}^3\dfrac{1}{\g_k}\left(\dfrac{\partial \Lambda _{j k}}{\partial \g_i}\,\g_iM_ i -
\dfrac{\partial \Lambda_{i k} }{\partial \g_ j}\,\g_jM_ j\right)
\eq
In the similar notations at $C_1=0$  the second bivector   is equal to
\bq\label{p-2}
P'=\alpha\,\left(
  \begin{array}{cc}
   \varkappa^2 \partial_\g\Lambda  & \Lambda \\ \\
    -\Lambda^\top &0
  \end{array}
\right)+\left(
  \begin{array}{cc}
   \partial_M\Pi & \Pi \\ \\
    -\Pi^\top & 0
  \end{array}
\right)+\left(
  \begin{array}{cc}
   0 & \Lambda' \\ \\
    -\Lambda'^\top & -\partial_\g\Lambda'
  \end{array}
\right)\,,\eq
where
\[
\alpha=\dfrac{2\varkappa^2(\beta_1M_1^2+\beta_2M_2^2)}{(\beta_1-\beta_2)\g_3^2}\,,\qquad
\beta_1=\varkappa^2\,a_1-1\,,\quad \beta_2=\varkappa^2\,a_2-1\,.
\]
Matrix $\Pi$ is equal to
\[
\Pi=\dfrac{2\varkappa^2}{(\beta_1-\beta_2)\g_3^2}
\left(\begin{smallmatrix}
\beta_1M_1(M_2\g_3-\g_2M_3)\quad&-\beta_1(\g_2M_2M_3+\g_3M_1^2)\quad & \beta_1\g_2(M_1^2+M_2^2)\\ \\
\beta_2(\g_1M_1M_3+\g_3M_2^2)\quad&-\beta_2M_2(M_1\g_3-\g_1M_3)\quad& -\beta_2\g_1(M_1^2+M_2^2)\\ \\
\beta_2\g_3M_2M_3&-\beta_1\g_3M_1M_3 & (\beta_1-\beta_2)\g_3M_1M_2
\end{smallmatrix} \right)\,,
\]
and antisymmetric matrix $\partial_M \Pi$ reads as
\[
\partial_M \Pi=\varkappa^2
                            \left(\begin{smallmatrix}
                              0 & M_3 & M_2 \\  \\
                              -M_3 & 0 & M_1 \\  \\
                              -M_2 & -M_1 & 0
                            \end{smallmatrix} \right)
                        +\dfrac{2\varkappa^2M_3}{(\beta_1-\beta_2)\g_3}
                           \left(\begin{smallmatrix}
                              0 & \beta_2\g_3 &-\beta_1\g_2 \\  \\
                              -\beta_2\g_3 & 0 &\beta_2\g_1 \\  \\
                              \beta_1\g_2 & - \beta_2\g_1& 0
 \end{smallmatrix} \right)\,.
\]
Instead of antisymmetric matrix $\Lambda$ (\ref{l1}) in the second bivector we have symmetric matrix
\[ \Lambda'=\left(
                           \begin{array}{ccc}
                             0 & -\g_3 & \g_2 \\
                            - \g_3 & 0 & \g_1 \\
                             \g_2 & \g_1 & -\dfrac{2\g_1\g_2}{\g_3} \\
                           \end{array}
                         \right)\,,
\]
whereas definition of $\partial_\g \Lambda'$ is completely similar to   (\ref{dl1})
\bq
\Bigl(\partial_\g \Lambda' \Bigr)_{ij}=\sum_{k=1}^3\dfrac{1}{\g_k}\left(\dfrac{\partial \Lambda' _{j k}}{\partial \g_i}\,\g_iM_ i -
\dfrac{\partial \Lambda'_{i k} }{\partial \g_ j}\,\g_jM_ j\right)\,.
\eq
At $\varkappa\to 0$ one get bi-hamiltonian structure for the Chaplygin system, which has been obtained in \cite{ts10a}.
We believe that  bivector $P'$   has  some  algebro-geometric justification, similar to compatible bivectors on $so(n)$ from \cite{bb02}.

\begin{rem}
Usually the second Poisson bivector $P'$  is the Lie derivative of $P$ along some polynomial Liouville vector field $X$
\[
P'=\mathcal L_X(P)\,,
\]
see \cite{ts07c,ts08,ts09,ts10a}.  For the Bogoyavlensky system we could not  find such Liouville  vector field.  So, we can not  say that  bivector $P'$ (\ref{p-2}) is  the 2-coboundary associated with the Liouville vector field  $X$ in the Poisson-Lichnerowicz cohomology defined by $P$.
\end{rem}

To sum up, using applicable ans\"{a}tze for the Liouville vector field $X$ we get a real relatively simple quadratic bivector (\ref{p-2}) and some  more complicated complex  bivectors. Modern computer software allows to do it on a personal computer wasting only few seconds. The application of this Poisson bivector will be given in the next section.

\section{Variables of separation and separation relations}
\setcounter{equation}{0}
The second step in the bi-hamiltonian method of separation of variables is calculation of canonical variables of separation $(q_1,\dots,q_n,p_1,\dots,p_n)$ and separation relations of the form
\begin{equation}
\label{seprelint}
\phi_i(q_i,p_i,H_1,\dots,H_n)=0\ ,\quad i=1,\dots,n\ ,
\qquad\mbox{with }\det\left[\frac{\partial \phi_i}{\partial H_j}\right]
\not=0\> .
\end{equation}
The reason for this definition is that the stationary
Hamilton-Jacobi equations for the Hamiltonians $H_i$ can be
collectively solved by the additively separated complete integral
\begin{equation}\label{eq:i0}
W(q_1,\dots,q_n;\alpha_1,\dots,\alpha_n)=
\sum_{i=1}^n W_i(q_i;\alpha_1,\dots,\alpha_n)\>,
\end{equation}
where  $W_i$ are found by quadratures as  solutions of ordinary
differential equations.

According to \cite{fp02,ts08,ts09},  separated coordinates $q_j$ are the eigenvalues of the control matrix $F$ defined by \[
P'{\mathbf{dH}}=P\bigl(F{\mathbf{dH}}\bigr).
\]
Its eigenvalues coincide with the Darboux-Nijenhuis coordinates (eigenvalues of the recursion operator) on the corresponding symplectic leaves. Using control matrix $F$ we can avoid the procedure of restriction of the bivectors $P$ and $P'$ on symplectic leaves, that is a necessary intermediate calculation for the construction of the recursion operator \cite{fp02}.

In our case for the  Poisson bivector $P'$ (\ref{p-2})  control matrix $F$ reads as
\bq
F=\dfrac{1}{2v}\left(
 \begin{array}{cc}
2u &1\\ \\
H_2& 2u
 \end{array}
\right)\,,
\eq
where
\[
 u= -(\beta_1M_1^2+\beta_2M_2^2)\,,\qquad v=-\varkappa^{-2}(\beta_1-\beta_2)p_3^2\,.
\]
The eigenvalues of this matrix $F$ are the required variables of separation $q_{1,2}$
\bq\label{sep-q}
q_1=\dfrac{u+\sqrt{H_2}}{v}\,,\qquad q_2=\dfrac{u-\sqrt{H_2}}{v}\,.
\eq
These variables has been introduced in  \cite{bog91} without any explanations and reasonable arguments. We reproduce this result in framework of the  generic method based on direct solution of the equations  (\ref{m-eq12}).
According to \cite{fp02},  eigenvectors of the control matrix $F$ form the St\"ackel matrix $S$
 \[
 F=S\,\left(
        \begin{array}{cc}
          q_1 & 0 \\
          0 & q_2 \\
        \end{array}
      \right)
 \,S^{-1}
 \]
whose entries  $S_{ij}$  depend only on a pair  $(q_i,p_i)$ of the canonical variables of separation. In our case matrix $S$ is equal to
\[
S=\left(
 \begin{array}{cc}
   1 & 1\\
  2\sqrt{H_2} & -2\sqrt{H_2} \\
 \end{array}
 \right)\,.
\]
It means  that we have non-St\"ackel integrable system with non-affine in $H_1$ or $H_2$ separated relations (\ref{seprelint}), similar to the generalized Chaplygin system \cite{ts10a} and the Kowalevski top \cite{ts10b}.

From the definitions of separation coordinates $q_{1,2}$ (\ref{sep-q}) and $H_2$ (\ref{bog-h}) we immediately obtain
\[
M_1=\sqrt{\dfrac{(1-q_1)(1-q_2)}{2\beta_1(q_1-q_2)}}\,H_2^{1/4}\,,\qquad
M_2=\sqrt{\dfrac{(1+q_1)(1+q_2)}{2\beta_2(q_2-q_1)}}\,H_2^{1/4}\,,\]
and
\[
\g_3=\dfrac{2\varkappa H_2^{1/4}}{\sqrt{2(\beta_1-\beta_2)(q_2-q_1)}}\,.
\]
Such as  $C_1=0$ and
\[
\dot{q}_k=\{H,q_k\}=-\dfrac{4\beta_1p_2M_1(1+q_k)}{\g_3}-\dfrac{4\beta_2\g_1M_2(1-q_k)}{\g_3}\,,
\]
we have
\ben
\g_1&=&\dfrac{\varkappa}{4(q_1-q_2)\sqrt{\beta_2(\beta_1-\beta_2)}}\dfrac{(1+q_2)\,\dot{q}_1-(1+q_1)\,\dot{q}_2}{\sqrt{(1+q_1)(1+q_2)}}\,, \nn\\ \nn\\
\g_2&=&\dfrac{\varkappa}{4(q_1-q_2)\sqrt{\beta_1(\beta_2-\beta_1)}}\dfrac{(1-q_2)\,\dot{q}_1-(1-q_1)\,\dot{q}_2}{\sqrt{(1-q_1)(1-q_2)}}\,,\nn\\
\nn\\
M_3&=&\dfrac{\dot{q}_1(1-q_2^2)+\dot{q}_2(1-q_1^2)}{4(q_1-q_2)\sqrt{-\beta_1\beta_2(1-q_1^2)(1-q_2^2)}}\,.\nn
\en
Substituting these expressions into the Hamiltonian  $H_1$ and Cazimir $C_2$ and solving the resulting equations
with respect to $\dot{q}_{1,2}$ one gets a pair of the Abel-Jacobi  equations
\bq\label{abel-eq}
\dfrac{\varkappa \dot{q}_1}{\sqrt{4(q_1^2-1)(\lambda_1q_1+\mu_1)}}=1\,,\qquad
\dfrac{\varkappa \dot{q}_2}{\sqrt{4(q_2^2-1)(\lambda_2q_2+\mu_2)}}=1\,.
\eq
where
\ben
\lambda_{1,2}&=&(\beta_1-\beta_2)\Bigl(C_2(\beta_1+\beta_2)+\varkappa^2(H_1\pm\sqrt{H_2})\Bigr)\,,\nn\\
\nn\\
\mu_{1,2}&=&(\beta_1-\beta_2)^2C_2+\varkappa^2(\beta_1+\beta_2)(H_1\pm\sqrt{H_2})\,.\nn
\en
Let us note that~(\ref{abel-eq}) are degenerate
Abel-Jacobi equations, i.\,e. each of them depends on a unique
variable  $q_1$ or $q_2$ only, and a two-dimensional Abel torus
splits into one-dimensional tori.

According to  \cite{tsbook}, the remaining separation variables  $p_{1,2}$   are equal to
\[
p_k=\dfrac{\dot{q}_k}{4(1-q_k^2)\bigl(\beta_1(1+q_k)+\beta_2(1-q_k)\bigr)}\,,\qquad \{q_i,p_k\}=\delta_{ik}\,,\quad i,k=1,2.
\]
They satisfy  to  the following  separated relations which directly  follow from the Abel-Jacobi equations (\ref{abel-eq}).
\bq\label{sep-rel}
\Phi_{1,2}(q,p)=4\varkappa^2(q^2-1)\Bigl(\beta_1(1+q)+\beta_2(1-q)\Bigr)^2p^2-\lambda_{1,2}q-\mu_{1,2}=0\,,
\eq
here $q=q_{1,2}$ and $p=p_{1,2}$.  At $\varkappa\to 0$ these equations coincide with the separated relations for the Chaplygin system, see \cite{ts10a}.

The third part of the Jacobi method consists of the construction of new integrable systems starting with known variables of separation and some other  separated relations. Namely, if we substitute our variables of separation $q_{1,2}$ and $p_{1,2}$ into the following deformation of  (\ref{sep-rel})
\bq\label{def-1}
\Phi^{(d)}(q,p)=\Phi_1\Phi_2-d_1 q-d_2=0,\qquad d_1,d_2\in\mathbb R,
\eq
and solve the resulting equations with respect to integrals of motion $H_{1,2}$, then we get rational generalization of the initial polynomial Hamilton function
\ben
H_1^{(d)}&=&H_1
-\dfrac{\varkappa^2}{16\beta_1^2\beta_2^2(\beta_1-\beta_2)(\varkappa^2M_1^2+\varkappa^2M_2^2+\g_3^2)^2}
\Bigl(\beta_1\varkappa^2(d_1-d_2)(\beta_1-\beta_2)M_1^2\Bigr.\nn\\
\nn\\
&-&\Bigl.\beta_2\varkappa^2(d_1+d_2)(\beta_1-\beta_2)M_2^2
+\bigl(\beta_1^2(d_1-d_2)+\beta_2^2(d_1+d_2)\bigr)\g_3^2\Bigr)
\,.
\nn
\en
If $d_1 =d_2(\beta_1-\beta_2)/(\beta_1+\beta_2)$ this Hamiltonian looks like
\[
H_1^{(d)}=H_1+\dfrac{2d_2\varkappa^2\beta_1\beta_2(\beta_1-\beta_2)}
{(\beta_1+\beta_2)(\varkappa^2M_1^2+\varkappa^2M_2^2+\g_3^2)}\,.
\]
At $\varkappa\to 0$ we obtain the Hamilton function for the generalized Chaplygin system studied in  \cite{ts10a}.
The main problem of this part of the Jacobi method  is how to get the Hamiltonian to be  interesting to physics.

\section{Conclusion}
Starting with the integrals of motion for the  Bogoyavlensky system on $so(4)$  we found  polynomial in momenta Poisson bivector $P'$, which are compatible with the canonical Poisson bivector $P$ on zero-level of the Casimir function $C_1$.
Then  in framework of the bi-hamiltonian geometry we   reproduce known separation  variables and separated relations. Some rational generalization of the Bogoyavlensky system is considered.

This example may  be useful for  creating a  general theory, which takes the  constructive answers  on the main open questions:
 \begin{itemize}
  \item how to get the  Poisson bivectors $P'$  on $so(n)$ compatible with $P$ ;
  \item how to describe all the natural Hamilton functions associated with  a given $P'$.
\end{itemize}
 Now we have some particular answers obtained  by direct tedious computations  only \cite{bb02,ts07c,ts08}.

The author wish to thank A.V. Tsiganov for formulation of the problem and  stimulating discussions.

\end{document}